\documentclass[conference]{IEEEtran}
\IEEEoverridecommandlockouts
\usepackage{cite}
\usepackage{amsmath,amssymb,amsfonts}
\usepackage{algorithmic}
\usepackage{graphicx}
\usepackage{textcomp}
\usepackage{longtable}

\usepackage[table]{xcolor}
\usepackage{hyperref}
\def\BibTeX{{\rm B\kern-.05em{\sc i\kern-.025em b}\kern-.08em
    T\kern-.1667em\lower.7ex\hbox{E}\kern-.125em}}
\begin{document}

\title{An approach to optimize inference of the DIART speaker diarization pipeline\\}

\author{\IEEEauthorblockN{1\textsuperscript{st} Roman Aperdannier}
\IEEEauthorblockA{\textit{Faculty of Business} \\
\textit{University of Applied Science}\\
Ansbach, Germany \\
aperdannier19472@hs-ansbach.de}
\and
\IEEEauthorblockN{2\textsuperscript{nd} Sigurd Schacht}
\IEEEauthorblockA{\textit{Faculty of Business} \\
\textit{University of Applied Science}\\
Ansbach, Germany \\
sigurd.schacht@hs-ansbach.de}
\and
\IEEEauthorblockN{3\textsuperscript{rd} Alexander Piazza}
\IEEEauthorblockA{\textit{Faculty of Business} \\
\textit{University of Applied Science}\\
Ansbach, Germany \\
alexander.piazza@hs-ansbach.de}

}

\maketitle

\begin{abstract}
Speaker diarization answers the question "who spoke when" for an audio file. In some diarization scenarios, low latency is required for transcription. Speaker diarization with low latency is referred to as online speaker diarization. The DIART pipeline is an online speaker diarization system. It consists of a segmentation and an embedding model. The embedding model has the largest share of the overall latency. The aim of this paper is to optimize the inference latency of the DIART pipeline. Different inference optimization methods such as knowledge distilation, pruning, quantization and layer fusion are applied to the embedding model of the pipeline. It turns out that knowledge distillation optimizes the latency, but has a negative effect on the accuracy. Quantization and layer fusion also have a positive influence on the latency without worsening the accuracy. Pruning, on the other hand, does not improve latency.
\end{abstract}

\begin{IEEEkeywords}
online speaker diarization, inference optimization, pruning, knowledge distillation, layer fusion, quantization
\end{IEEEkeywords}

\section{Introduction}
With speaker diarization, audio segments are assigned to their corresponding speakers. This information is essential for a complete audio transcription. This is why speaker diarization in combination with automatic speech recognition (ASR) is used in many transcription scenarios such as online meetings, earnings reports, court proceedings, interviews, etc. \cite{park_review_2021}. In some of these scenarios, low transcription latency is necessary. For example, the transcription of an earnings report in near real time can provide a competitive advantage.  As automated share purchases or sales can be executed immediately on the basis of the information obtained \cite{de_castro_mt5b3_2021}. Speaker diarization with a correspondingly low latency is referred to as online speaker diarization.\\

In the paper of Aperdannier \cite{aperdannier_systematic_2024}, various online speaker diarization systems are evaluated in terms of their latency. A speaker diarization pipeline from the DIART framework \cite{coria_overlap-aware_2021} turns out to be the fastest system. The pipeline consists of the speaker embedding model \textit{pyannote/embedding} and the segmentation model \textit{pyannote/segmentation}. The paper also shows that the segmentation model has a comparably low influence on the latency of the overall system.\\

The aim of this work is to optimize the latency of the DIART pipeline. The focus is on the embedding model, as this is the largest factor in the overall latency \cite{aperdannier_systematic_2024}. Techniques such as knowledge distilation, pruning, quantization, layer fusion and export to Open Neural Network Exchange (ONNX) \cite{ambi_onnx_2021} format are used for inference optimization of the model. The research question of this paper can be formulated as follows. Which methods can best optimize the inference speed of the speaker embedding model \textit{pyannote/embedding}? \\

The paper is structured as follows. In the next section, the inference optimization methods and the scientific background  are explained. The methodology of the paper is then discussed in more detail. Afterwards the research method, data collection and machine learning algorithms are presented. The experimental setup with model training and evaluation setup is then described. Finally, the results are shown and discussed.

\section{Background and Related Work}
As described in the introduction, the aim of this work is to optimize the inference speed of the speaker diarization pipeline of the DIART framework. There are many common methods for the inference optimization of machine learning models. The methods used here are explained in more detail in the following sections.

\subsection{Knowledge Distillation}
Knowledge distillation is one of the methods for inference optimization. A complex teacher model is used to train a simpler student model. This teacher-student relationship can be established, for example, by an additional loss function that minimizes the distance between the predictions of the teacher model and the predictions of the student model \cite{romero_fitnets_2015}. The result is a student model that achieves similar performance to the teacher model, but with fewer computational and memory resources \cite{hinton_distilling_2015}. In principle, knowledge distillation can be applied to all learning systems. However, knowledge distillation is particularly advantageous for deep neural networks such as convolutional neural networks (CNNs) \cite{aghli_combining_2021} \cite{ahmed_compact_2021} or transformers \cite{jiang_knowledge_2021} \cite{lin_knowledge_2022}.

\subsection{Pruning} \label{pruning}
The main goal of pruning is to reduce the size of a neural network. This is achieved by reducing the connections in the network by deleting neurons or zeroing out weights \cite{molchanov_pruning_2017}. Pruning methods can be divided into structured and unstructured \cite{molchanov_pruning_2017}. Structured pruning methods \cite{lin_runtime_2017} \cite{hu_network_2016} remove entire channels or layers. Unstructured pruning methods \cite{han_learning_2015} \cite{molchanov_variational_2017}, on the other hand, deactivate individual neurons by setting their weight to zero. However, this means that unstructured pruning does not automatically make the model smaller or faster \cite{liu_rethinking_2019}. This requires additional libraries that can generate performance gains from the zeroed neurons \cite{liu_rethinking_2019}.

\subsection{Quantization} \label{quant}
Quantization is another method for optimizing neural networks. Quantization reduces the number of bits used to represent weights and activations in a neural network. The \textit{int8} quantization is most commonly used, as a good balance between performance gain and accuracy loss, which can be observed \cite{jacob_quantization_2018} \cite{lin_fixed_2016} \cite{wu_quantized_2016}. Common deep learning frameworks such as PyTorch or Tensorflow have therefore also started to implement \textit{int8} quantizations \cite{pytorch_contributors_quantization_2023} \cite{tensorflow_post-training_2024}.

\subsection{Layer Fusion}
Layer fusion combines several successive layers of a model into a single layer. This can reduce the number of processes that need to be carried out during inference \cite{chen_multi-layer_2018} \cite{pytorch_contributors_fuse_modules_2023}. Layer fusion is mainly applied to deep neural networks such as CNNs \cite{chen_multi-layer_2018} and transformers \cite{kao_dnnfuser_2022} \cite{liu_understanding_2021}.

\subsection{ONNX Format} \label{onnx}
Another way to improve the inference speed of a model can be to use a different runtime. Some papers and block contributions evaluate the inference speedup by converting a model from PyTorch to ONNX \cite{ambi_onnx_2021} \cite{chaigneau_boost_2022} \cite{zhou_exploring_2022}. ONNX is an intermediate format for machine learning models \cite{onnx_onnx_2024}. In addition, the ONNX runtimes offer the possibility of hardware optimizations \cite{onnx_onnx_2024}.

\section{Methodology}
This section provides an overview of the research method, data collection and used machine learning algorithms.

\subsection{Research Method}
The aim of this paper is to apply different inference optimization methods on the model \textit{pyannote/embedding} and then measure their effect on the inference speed. For this purpose, a structured experiment is conducted. At the beginning, a reduced \textit{pyannote/embedding} model is created and trained with knowledge distillation. There is no PyTorch layer fusion for the architecture of the \textit{pyannote/embedding} model. Therefore, the further inference optimizations are then applied to the reduced model to ensure comparability. Subsequently, the resulting optimized embedding models are used to run the speaker diarization pipeline of the DIART framework with a uniform dataset on uniform hardware. Diarization error (DER) \cite{fiscus_rich_2006} and latency are measured from audio input to speaker label output. The DIART pipeline with the unoptimized \textit{pyannote/embedding} model serves as the baseline. Finally, the measurement results are classified and discussed.

\subsection{Data Collection} \label{data_collection}
The dataset AMI corpus \cite{carletta_ami_2005} is used to train the reduced \textit{pyannote/embedding} model. Pyannote provides a Githup repository \cite{landini_bayesian_2020} with a setup for the AMI corpus. The dataset can then be used to train a Pyannote \cite{bredin_pyannoteaudio_2023} compatible model via corresponding shell scripts. No further data is required for the other inference optimization methods.
As a testset for the subsequent evaluation of the optimized models, a subset of the Voxconverse \cite{chung_spot_2020} testset is used in this work. The subset consists of the first four audio files aepyx.wav, aggyz.wav, aiqwk.wav and auzuru.wav. These contain a total of around 20 minutes of audio recordings.

\subsection{Machine Learning Models}
\begin{figure} [h]
    \centering
    \includegraphics[width=0.47\linewidth]{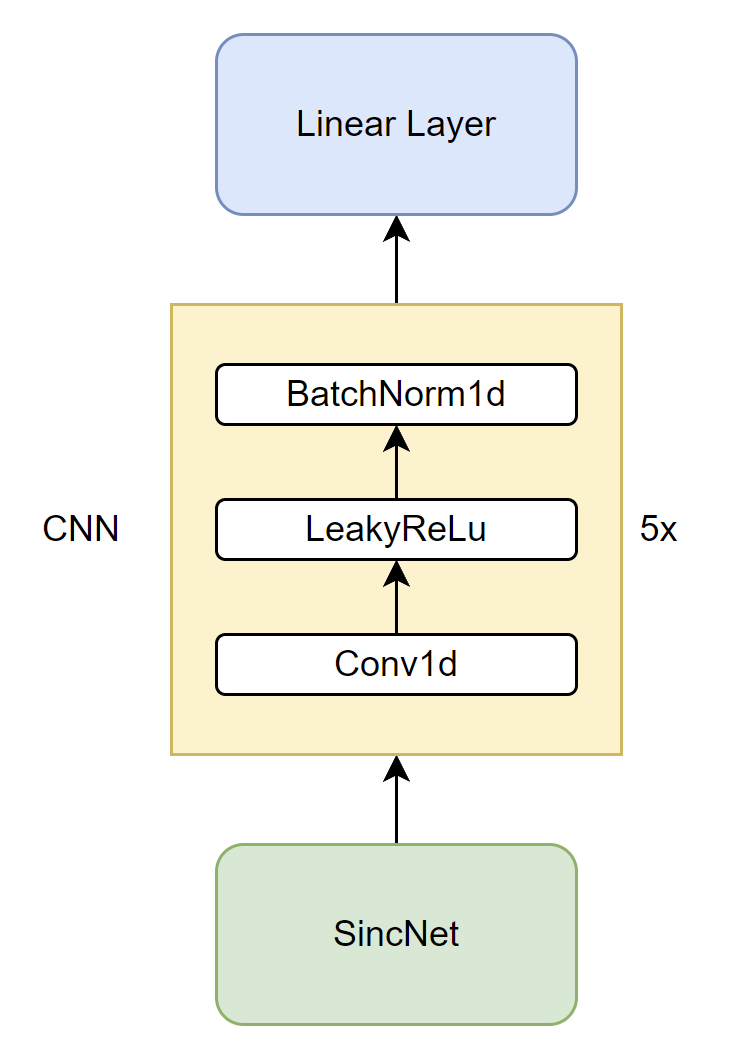}
    \caption{Pyannote embedding structure}
    \label{fig:pyannote_emb}
\end{figure}

As described above, the pipeline of the DIART framework with embedding model \textit{pyannote/embedding} and the segmentation model \textit{pyannote/segmentation} is used as the baseline. 
The structure of the \textit{pyannote/embedding} model \cite{bredin_pyannoteaudio_2020} can be seen in figure \ref{fig:pyannote_emb}. SincNet is a Convolutional Neural Network (CNN)-based embedder that receives audio features as input and converts them into tensors. The tensors are then further processed by a CNN with five layers. Each of the five layers consists of a convolutional layer, a LeakyRelu and a subsequent BatchNorm. The output of the CNN is then passed to the linear layer, which generates the final audio embeddings. \\

\begin{figure} [h]
    \centering
    \includegraphics[width=0.47\linewidth]{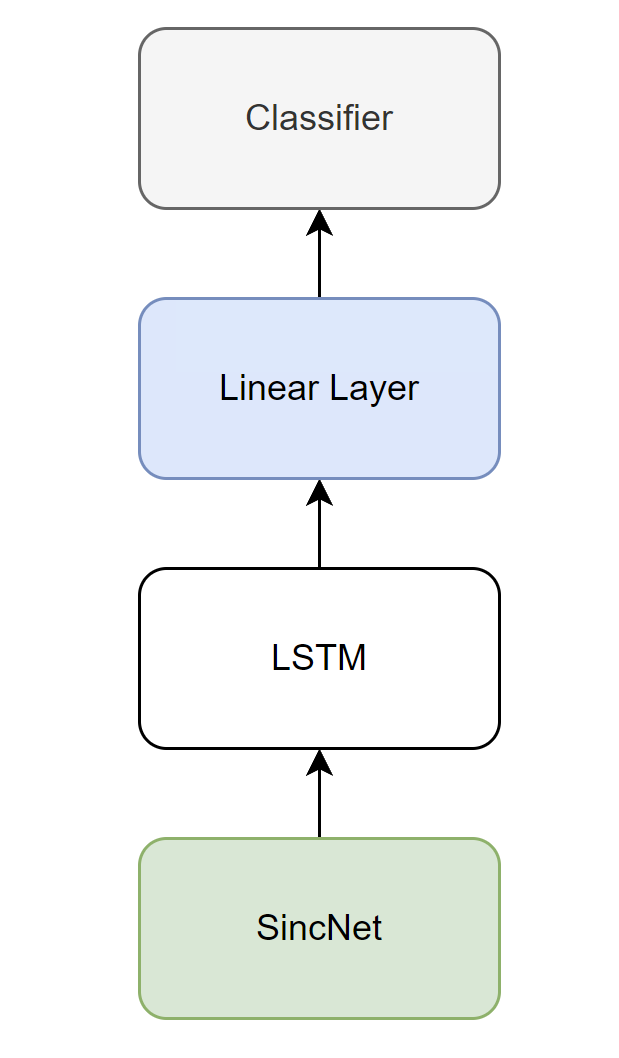}
    \caption{Pyannote segmentation structure}
    \label{fig:pyannote_seg}
\end{figure}

The architecture of the \textit{pyannote/segmentation} model \cite{bredin_pyannoteaudio_2020} can be seen in figure \ref{fig:pyannote_seg}. SincNet is the same network as in \textit{pyannote/embedding}. The Long Short-Term Memory (LSTM) is then used as a classifier to return the corresponding audio class (e.g. No Speaker, Speaker 1, Speaker 2 ...).

A reduced \textit{pyannote/embedding} model is used for knowledge distillation. Its architecture is described in more detail in \ref{model_arch} Model architecture.
\newpage
\section{Experimental Setup}
This chapter describes the structure of the experiment. First, the architecture of the reduced \textit{pyannote/embedding} model is described. Then the training with knowledge distillation is presented. Finally, the implementation of the inference optimizations is described.

\subsection{Model Architecture} \label{model_arch}
\begin{figure}[h]
    \centering
    \includegraphics[width=0.5\linewidth]{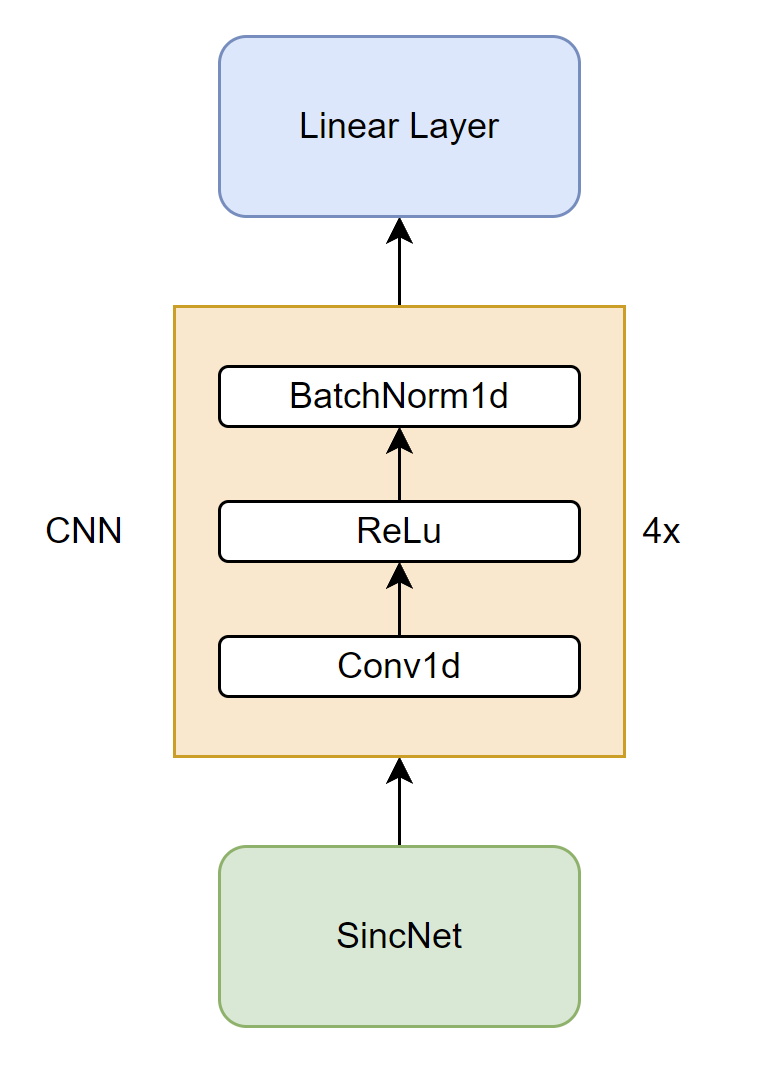}
    \caption{Pyannote embedding reduced structure}
    \label{fig:pyannote_emb_reduced}
\end{figure}

The model for knowledge distillation has a similar structure to the \textit{pyannote/embedding} model. As can be seen in figure \ref{fig:pyannote_emb_reduced}, only one of the CNN levels has been removed. In addition, the activation function LeakyReLu has been replaced by ReLu. The reason for this is that ReLu enables subsequent layer fusion with PyTorch. PyTorch has not yet implemented layer fusion for the LeakyReLu activation function \cite{pytorch_contributors_fuse_2024}. The model is referred to as \textit{pyannote/embedding\_reduced} in the following sections.

\subsection{Model Training - Knowledge Distillation}
Pyannote provides a pipeline for training Pyannote compatible models \cite{bredin_pyannoteaudio_2023}. The pipeline consists of a model and the corresponding task. In the case of this work, the model is \textit{pyannote/embedding\_reduced} with a custom task. The standard tasks of the Pyannote training pipeline do not provide the option of integrating knowledge distillation into the training process. Therefore, a custom task is created for the training in this paper. This is based on the Pyannote task \textit{SupervisedRepresentationLearningWithArcFace}. The loss function of the standard task is adjusted by including the distance to the teacher embedding. The \textit{pyannote/embedding} model is used as the teacher model. In addition the distance to the teacher embedding is multiplied by a factor $\lambda$. This implementation is based on the paper by Romero et al. \cite{romero_fitnets_2015}. In order to determine a good value for the factor, training sessions were carried out with 50 epochs and different $\lambda$ values. The corresponding DER results on the testset can be seen in table \ref{tab:teacherfactor} teacher factor. $\lambda = 1000$ turns out to be a good value for knowledge distillation.

\begin{table} [h]
    \centering
    \caption{teacher factor $\lambda$ evaluation}
    \begin{tabular}{|l|l|}
         \hline
         \rowcolor{lightgray!70} \textbf{factor $\lambda$}& \textbf{DER}\\
         \hline
         0& 60.161583\\
         \hline
         1& 61.24425\\
         \hline
         10& 57.82841\\
         \hline
         100& 52.05711\\
         \hline
         \rowcolor{orange!30} 1000& 50.169207\\
         \hline
         10 000& 52.767206\\
         \hline
         100 000& 53.064924\\
         \hline
    \end{tabular}
    \label{tab:teacherfactor}
\end{table}

In addition, the Pyannote training pipeline does not currently offer the option of using early stopping. However, to avoid overfitting and underfitting, training is then carried out with $\lambda = 1000$ and a checkpoint every 20 epochs. The checkpoints are each evaluated with the testset mentioned in \ref{data_collection} Data Collection. Table \ref{tab:train_val} training validation shows the DER results of the checkpoints.

\begin{table}[h]
    \centering
    \caption{training validation}
    \begin{tabular}{|l|l|}
         \hline
         \rowcolor{lightgray!70} \textbf{epoch}& \textbf{DER}\\
         \hline
         19& 53.0094382\\
         \hline
         39& 52.933049\\
         \hline
         59& 52.875565\\
         \hline
         \rowcolor{orange!30} 79& 49.726088\\
         \hline
         99& 52.561401\\
         \hline
         119& 52.543265\\
         \hline
         139& 52.594601\\
         \hline
    \end{tabular}
    \label{tab:train_val}
\end{table}

It can be seen that the DER decreases until epoch 79. The DER then increases again and stabilizes around 52.5\%. It is striking that the DER for epoch 59 in table \ref{tab:train_val} is almost 2.5\% higher than for the training with 50 epochs in table \ref{tab:teacherfactor}. This is presumably due to the fact that Pyannote slices the data differently for each training. However, no further analysis of this issue is carried out in this paper. 
For the further optimizations in the following sections, the model of the checkpoint at epoch 79 is used. This contains a \textit{pyannote/embedding\_reduced} that was trained with a teacher factor $\lambda = 1000$ and correspondingly 79 epochs.

\subsection{Inference Optimization}
The following sections describe the implementations of inference optimization methods.
\\
\subsubsection{Pruning} \label{pruning}
The prune-module of the PyTorch library is used for pruning \cite{pytorch_contributors_torchnn_2023}. On the one hand, structured pruning is performed with the function 
\begin{itemize}
    \item \textit{ln\_structured(module, name, amount=1, n=2, dim=0)}.
\end{itemize}
On the other hand, unstructured pruning is carried out with the function 
\begin{itemize}
    \item \textit{global\_unstructured(modules, L1Unstructured, amount =0.3)}.
\end{itemize}
As parameters the default values are used. Only the weights of the convolutional and linear modules of the \textit{pyannote/embedding\_reduced} model are pruned.
\\
\subsubsection{Layer Fusion}
Layer fusion is performed with the \textit{fuse\_modules} function from \cite{pytorch_contributors_fuse_2024}. The function offers the possibility to combine certain sequences of modules into one module. In this work, all \textit{conv1d-relu-sequences} are fused for the \textit{pyannote/embedding\_reduced} model.
\\
\subsubsection{Quantization}
As described in the chapter \ref{quant} Quantization, a good ratio of accuracy loss and performance gain can usually be observed for \textit{int8} quantization \cite{jacob_quantization_2018} \cite{lin_fixed_2016} \cite{wu_quantized_2016}. For this reason, an \textit{int8} quantization with the PyTorch Library \textit{torch.quantization} \cite{pytorch_contributors_quantization_2023} is performed in this work. Some sources also show that activation quantization has a higher sensitivity in terms of accuracy than weight quantization \cite{park_weighted-entropy-based_2017} \cite{yao_zeroquant-v2_2023}. For this reason, a pure weight quantization with the dynamic quantization of Pytroch is carried out in this work.
\\
\subsubsection{ONNX}
The corresponding PyTorch library \cite{pytorch_contributors_torchonnx_2023} is also utilized to convert the \textit{pyannote/embedding\_reduced} into the ONNX format. The \textit{pyannote/embedding\_reduced} model is transferred to ONNX format using the \textit{onnx.export} function. The resulting onnx model can then be loaded into the DIART pipeline. The DIART framework provides a corresponding loader for this \cite{coria_overlap-aware_2021}.

\subsection{Evaluation Setup}
As described in \ref{data_collection} Data collection, a subset of the Voxconverse testset is used for the evaluation. The DIART pipeline is evaluated with the \textit{pyannote/segmentation} model and the corresponding optimized model. The evaluation is carried out on an Intel(R) Xeon(R) Gold 5215 CPU @ 2.50GHz. The DIART pipeline is called with the following default values:
\begin{itemize}
    \item step=0.25
    \item latency=3.0
    \item tau\_active=0.555
    \item rho\_update=0.422
    \item delta\_new=1.517
\end{itemize}

The Python library time is then used to measure the following time span for each chunk/step:
\begin{equation}
\begin{split}
latency = & t_{2}(\text{time of the speaker label output}) \\
- & t_{1}(\text{time of the audio chunk input})
\end{split}
\end{equation}

For the time measurement the Python function \textit{time.perf\_counter} is used. This function uses a performance counter with the highest possible resolution on the test system \cite{python_documentation_time_2024}.

The DIART framework already offers a way to measure the latency for processing a chunk. Here, only the used time function \textit{time.monotonic} is exchanged for \textit{time.pref\_counter}. In this work, the mean and the standard deviation of the corresponding latency values are reported.

\section{Results}
\subsection{Latency}
The results of the evaluation are shown in table \ref{table:eval_infernce}. It can be seen that knowledge distillation results in an improvement in latency of just under 10\%. The two pruning methods do not lead to any further improvement in latency. However, layer fusion and quantization on the other hand lead to a further improvement of just under 7\%. Converting the model to onnx format leads to a significant deterioration in latency.

\subsection{DER}
The diarization error increases by 5\% in absolute terms due to the knowledge distillation. The two pruning methods and quantization lead to a further slight deterioration in the DER. However, quantization only influences the DER in the third descendant position. Layer fusion and conversion to the ONNX format have no influence on the DER.

\subsection{Model Size}
The model size represents the memory size of the associated \textit{state\_dict} \cite{pytorch_contributors_what_2023}. A reduction from 16.62 MB to 13.61 MB can be observed in the model size due to knowledge distillation. The \textit{int8} quantization for a further reduction of the model size to 9.22 MB. The other inference optimization methods have no influence on the model size.

\begin{table*}[t]
  \centering
    \caption{Evaluation results - Inference Optimization}
      \begin{tabular}{|p{5.5cm}|p{2cm}|p{2cm}|p{2cm}|p{2cm}|}
        \hline
        \rowcolor{lightgray!70} \textbf{Model} & \textbf{DER} & \textbf{Pipeline latency mean in s}& \textbf{Latency in \%} & \textbf{Model size in MB}\\
        \hline
        Baseline & 44.859316 &	0.065127 &	100.0 &	16.62 \\ 
        \hline
        \rowcolor{orange!30}knowledge distillation & 49.726088	& 0.058739	& 90.2	& 13.61 \\
        \hline
        knowledge distillation + structured pruning & 49.793409	& 0.058558 &	89.9	&13.61 \\
        \hline
        knowledge distillation + unstructured pruning & 50.20256 &	0.059035 &	90.6 &	13.61 \\
        \hline
        \rowcolor{orange!30}knowledge distillation + layer fusion & 49.726088	& 0.054422 &	83.6 &	13.61\\
        \hline
        \rowcolor{orange!30}knowledge distillation + int8 quantization & 49.727625 &	0.054007&	82.9 &	9.22\\
        \hline
        knowledge distillation + ONNX & 49.726088 &	0.090893	&139.6&	13.62\\
        \hline
      \end{tabular}
  \label{table:eval_infernce}
\end{table*}

\section{Discussion}
\subsection{Knowledge Distillation}
As described in \ref{model_arch} Model architecture, with knowledge distillation a model that has one convolution level less than the baseline model is trained. This reduces the number of parameters from 4.35 million to 3.56 million, which corresponds to a percentage reduction of approximately 18\%. The same percentage reduction can be observed for the model size. The reduction in model size also explains a corresponding reduction in latency.  The DER increases by just 5\% in absolute terms as a result of knowledge distillation. It is possible that the DER can be brought closer to the accuracy of the baseline system through more extensive training with additional datasets.\\

\subsection{Structured Pruning}
The structured pruning has no effect on the latency. To check whether pruning has a general effect on the model, the sparsity of the pruned modules was also tracked and the \textit{is\_pruned} function \cite{pytorch_contributors_torchnn_2023} was applied to the pruned modules. Sparsity is a value for the proportion of zeroed parameters in the pruned layers \cite{hoefler_sparsity_2021}. Structured pruning increases the average sparsity of the model from 0.0 to 0.60. In addition, the \textit{is\_pruned} function confirms that all modules have been pruned. However, the number of parameters remains constant at 3.56 million despite pruning. From this it can be concluded that no layers or entire neurons are removed by structured pruning in PyTorch and only some parameters are zeroed.

\subsection{Unstructured Pruning}
The unstructured pruning also has no effect on the latency. The sparsity of the model increases from 0.0 to 32.63. The pruning can also be confirmed by the \textit{is\_pruned} function. The parameter quantity also remains constant at 3.56 million, which means that only parameters are zeroed in unstructured pruning.\\  
As the average sparsity for unstructured pruning is relatively high at 32.63, an attempt was made to save the weights of the pruned models as sparse tensors in COO format \cite{pytorch_contributors_torchsparse_2023} in order to reduce the memory requirement. However, the memory requirement for the model in sparse format is 58.18 MB, which is significantly higher than the 13.61 MB in the classic dense format. One explanation for this is the following. \\

A coo-sparse tensor only stores the non-zero elements (nze) and the indices of these elements as \textit{int64}, which has a size of 8 Byte. The memory requirements of a coo-sparse tensor can therefore be mapped as follows \cite{pytorch_contributors_torchsparse_2023}:

\begin{equation}\label{eq:2}
\begin{split}
ndim * 8 * count(nze) + count(nze) * element_{size}
\end{split}
\end{equation}

The memory requirement for a classic dense tensor is \cite{pytorch_contributors_torchsparse_2023}:
\begin{equation}\label{eq:3}
\begin{split}
n * m * element_{size}
\end{split}
\end{equation}

For the PyTorch pruning of a tensor, the number of non-zero elements can be calculated by \(nze = n * m * (1-amount)\).The value of \(amount\) is the corresponding parameter of the pruning function. The resulting sparsity corresponds to the \(amount\) parameter with minor deviations. \\

From (\ref{eq:2}) and (\ref{eq:3}) it follows that the memory requirement of the coo-sparse tensor is only less than the memory requirement of the corresponding dense tensor if:
\begin{equation}
\begin{split}
amount > 1 - (element_{size}/ndim*8+ element_{size})
\end{split}
\end{equation}

For the weight tensor with \(ndim=3\) and \(element_{size}=4\) (size of \textit{float32}) of a convolution layer of the model \textit{pyannote/embedding\_reduced}, this is the case if \(amount > 0.858\). In case of this work an unstructured pruning with \(amount=0.30\) was carried out as it can be seen in \ref{pruning} Pruning. Therefore the pruned model only achieves a sparsity of approximately 32.63. This explains the significantly higher memory requirement in sparse format.

\subsection{Layer Fusion}
Layer fusion leads to a further reduction in latency of 7\%. The number of model parameters remains constant at 3.56 million due to the layer fusion, as the layers are only combined but not reduced. This means that the speedup cannot be explained by a smaller model. However, layer fusion reduces the overhead and fewer separate operations need to be executed sequentially \cite{chen_multi-layer_2018} \cite{pytorch_contributors_fuse_modules_2023}. This leads to a further speedup while maintaining the same accuracy.

\subsection{Quantization}
The int8 quatization transfers the data type of all weights of the model from \textit{float32} to \textit{int8}. This reduces the memory requirement per weight from 4 Bytes to 1 Byte \cite{pytorch_contributors_quantization_2023}. This explains the correspondingly reduced memory requirement of the quantized model. The activations remain unaffected by the applied quantization. 
The reduced bit size also reduces the effort required for the corresponding computing operations on the hardware \cite{park_weighted-entropy-based_2017}. As a result, quantization usually leads to an additional inference speedup, which in the case of this work is approximately 7\%.

\subsection{ONNX}
Converting the model to the ONNX format has only a marginal impact on the memory requirements. The reason for this is that the weights are adopted and only the metadata of the model is adapted \cite{pytorch_contributors_torchonnx_2023}.
The ONNX format makes it possible to use hardware optimizations and thus increase the inference speed \cite{onnx_onnx_2024}. In the case of this work, the conversion to the ONNX format degraded the inference speed of the DIART pipeline by almost 40\%. One explanation for this could be that the DIART pipeline with the ONNX model of the PyTorch \textit{onnx.export} function cannot use the corresponding hardware optimizations. This issue needs to be analyzed in detail in future research.

\section{Conclusion}
In this work, various inference optimization methods were applied to the \textit{pyannote/embedding} model. The resulting optimized embedding models were then evaluated in the DIART pipeline. \\

By training a reduced \textit{pyannote/embedding} model with knowledge distillation, the inference speed could be increased and the memory requirements reduced. However, the knowledge distillation leads to a loss of accuracy. The structured and unstructured pruning has no significant influence on latency, memory requirements and accuracy. Layer fusion leads to a further improvement in latency while memory requirements and accuracy remain the same. An improvement in latency and memory requirements can also be observed with int8 quantization. Accuracy is only marginally affected by quantization. As a further optimization method, the \textit{pyannote/embedding\_reduced} model was converted to the ONNX format. The ONNX format leads to a significant deterioration in latency while memory requirements and accuracy remain the same. Thus, the research question of this thesis can be answered as follows. The inference optimizations knowledge distillation, layer fusion and quantization have the greatest positive influence on the inference speed of the \textit{pyannote/embedding} model. \\

When the model was converted to the ONNX format, an improvement in latency was expected as described in \ref{onnx} ONNX. It would be interesting for future work to analyze the reason why the latency became higher instead. Future work could also carry out an optimized training with more training data for knowledge distillation. This could bring the accuracy closer to the accuracy of the baseline model. Thereby an increase of the cost-benefit ratio for the inference optimization methods could be obtained. \\

In summary, it can be stated that classical inference optimization methods such as knowledge distillation, pruning, layer fusion and quatization can be successfully applied to the audio embedding model of the DIART pipeline.

\bibliographystyle{IEEEtran}
\bibliography{references}

\end{document}